# Nash Convergence of Gradient Dynamics in General-Sum Games

**Satinder Singh**
AT&T Labs
Florham Park, NJ 07932
baveja@research.att.com

**Michael Kearns**
AT&T Labs
Florham Park, NJ 07932
mkearns@research.att.com

**Yishay Mansour**
Tel Aviv University
Tel Aviv, Israel
mansour@math.tau.ac.il

## Abstract

Multi-agent games are becoming an increasingly prevalent formalism for the study of electronic commerce and auctions. The speed at which transactions can take place and the growing complexity of electronic marketplaces makes the study of computationally simple agents an appealing direction. In this work, we analyze the behavior of agents that incrementally adapt their strategy through *gradient ascent* on expected payoff, in the simple setting of two-player, two-action, iterated general-sum games, and present a surprising result. We show that either the agents will converge to a Nash equilibrium, or if the strategies themselves do not converge, then their average payoffs will nevertheless converge to the payoffs of a Nash equilibrium.

## 1 Introduction

It is widely expected that in the near future, software agents will act on behalf of humans in many electronic marketplaces based on auctions, barter, and other forms of trading. This makes multi-agent game theory (Owen, 1995) increasingly relevant to the emerging electronic economy. There are many different formalisms within game theory that model interaction between competing agents. Our interest is in *iterated games*, which model situations where a set of agents or players repeatedly interact with each other in the same game. There is a long and illustrious history of research in iterated games, such as the study of (and even competitions in) solving iterated prisoner's dilemma (Owen, 1995). Of particular interest in iterated games is the possibility of the players *adapting* their strategy based on the history of interaction with the other players.

Many different algorithms for adaptive play in iterated games have been proposed and analyzed. For example, in *fictitious play*, each player maintains a model of the mixed strategy of the other players based on the empirical play so far, and always plays the best response to this model at each iteration (Owen, 1995). While it is known that the time averages of the strategies played form a Nash equilibrium, the strategies themselves do not converge to Nash, nor are the averaged payoffs to the players guaranteed to be Nash. Kalai and Lehrer (1993) proposed a Bayesian strategy for players in a repeated game that requires the players to have "informed priors", and showed that under this condition play converges to a Nash equilibrium. A series of recent results has shown that the informed prior condition is actually quite restrictive, limiting the applicability of this result.

These seminal results implicitly assume that unbounded computation is allowed at each step. In contrast, we envision a future in which agents may maintain complex parametric representations of either their own strategy or their opponents, and in which full Bayesian updating or computation of best responses is computationally intractable. In other words, as in the rest of artificial intelligence and machine learning, in order to efficiently act in a complex environment, agents will adopt both *representational* and *computational* restrictions on their behavior in iterated games and other settings (e.g., Papadimitrou and Yannakakis (1994) and Freund et al. (1995)).

Perhaps the most common type of algorithm within machine learning are those that proceed by gradient ascent or descent (or other local methods) on some appropriate objective function. In this paper we study the behavior of players adapting by gradient ascent in expected payoff in two-person, two-action, general-sum iterated games. Thus, here we study a specific and very simple adaptive strategy in a setting in which a general mixed strategy is easy to represent. Such a study is a prerequisite to an understanding of gradient methods on rich, parametric strategy representa-tions. While it is known from the game theory literature that the strategies computed by gradient ascent in two-person iterated games need not converge, we present a new and perhaps surprising result here. We prove that although the *strategies* of the two players may not always converge, their *average payoffs* always do converge to the expected payoffs of some Nash equilibrium. Thus, the dynamics of gradient ascent ensure that the average payoffs to two players adopting this simple strategy is the same as the payoff they would achieve by adopting arbitrarily complex strategies.

In the remaining sections, we define our problem and the gradient ascent algorithm, and show that the behavior of players adapting via gradient ascent can be modeled as an affine dynamical system. Many properties of this dynamical system are known from control theory literature, and have been applied before to the somewhat different setting of evolutionary game theory (Weibull, 1997). Our main technical contribution is a new and detailed geometric analysis of these dynamics in the setting of classical game theory, and particularly of the effects of the boundary conditions imposed by game theory on those dynamics (in contrast, evolutionary game theory explicitly and artificially prevents the dynamics from reaching the boundaries).

## 2 Problem Definition and Notation

A two-player, two-action, general-sum game is defined by a pair of matrices

$$R = \begin{bmatrix} r_{11} & r_{12} \\ r_{21} & r_{22} \end{bmatrix} \text{ and } C = \begin{bmatrix} c_{11} & c_{12} \\ c_{21} & c_{22} \end{bmatrix}$$

specifying the payoffs for the *row* player (player 1) and the *column* player (player 2), respectively. If the row player chooses action $i \in \{1,2\}$ and the column player chooses action $j \in \{1,2\}$ the payoff to the row player is $r_{ij}$ and the payoff to the column player is $c_{ij}$. Two cases of special interest are that of *zero-sum* games, in which the payoff of the column player and the payoff of the row player always sums to zero ($r_{ij} + c_{ij} = 0$ for $i,j \in \{1,2\}$), and that of *team* games, in which both players always get the same payoff ($r_{ij} = c_{ij}$ for $i,j \in \{1,2\}$).

The players can choose actions stochastically, in which case they are said to be following a *mixed* strategy. Let $0 \leq \alpha \leq 1$ denote the probability of the row player picking action 1 and let $0 \leq \beta \leq 1$ denote the probability of the column player picking action 1. Then $V_r(\alpha, \beta)$, the *value* or expected payoff of the strategy pair $(\alpha, \beta)$ to the row player, is

$$\begin{aligned}V_r(\alpha, \beta) &= r_{11}(\alpha\beta) + r_{22}((1-\alpha)(1-\beta)) \\ &\quad + r_{12}(\alpha(1-\beta)) + r_{21}((1-\alpha)\beta) \end{aligned} \quad (1)$$

and $V_c(\alpha, \beta)$, the value of the strategy pair $(\alpha, \beta)$ to the column player, is

$$\begin{aligned}V_c(\alpha, \beta) &= c_{11}(\alpha\beta) + c_{22}((1-\alpha)(1-\beta)) \\ &\quad + c_{12}(\alpha(1-\beta)) + c_{21}((1-\alpha)\beta). \end{aligned} \quad (2)$$

The strategy pair $(\alpha, \beta)$ is said to be a *Nash equilibrium* (or Nash pair) if (i) for any mixed strategy $\alpha'$, $V_r(\alpha', \beta) \leq V_r(\alpha, \beta)$, and (ii) for any mixed strategy $\beta'$, $V_c(\alpha, \beta') \leq V_c(\alpha, \beta)$. In other words, as long as one player plays their half of the Nash pair, the other player has no incentive to change their half of the Nash pair. It is well-known that every game has at least one Nash pair in the space of mixed (but not necessarily pure) strategies.

## 3 Gradient Ascent for Iterated Games

One can view the strategy pair $(\alpha, \beta)$ as a point in $\mathbb{R}^2$ constrained to lie in the unit square. The functions $V_r(\alpha, \beta)$ and $V_c(\alpha, \beta)$ then define two value surfaces over the unit square for the row and column players respectively. For any given strategy pair, $(\alpha, \beta)$, one can compute a gradient for the row player from the $V_r$-value surface and for the column player from the $V_c$-value surface as follows. Letting $u = (r_{11} + r_{22}) - (r_{21} + r_{12})$ and let $u' = (c_{11} + c_{22}) - (c_{21} + c_{12})$, we have

$$\frac{\partial V_r(\alpha, \beta)}{\partial \alpha} = \beta u - (r_{22} - r_{12}) \quad (3)$$

$$\frac{\partial V_c(\alpha, \beta)}{\partial \beta} = \alpha u' - (c_{22} - c_{21}). \quad (4)$$

In the gradient ascent algorithm, each player repeatedly adjusts their half of the current strategy pair in the direction of their current gradient with some step size $\eta$:

$$\begin{aligned}\alpha_{k+1} &= \alpha_k + \eta \frac{\partial V_r(\alpha_k, \beta_k)}{\partial \alpha} \\ \beta_{k+1} &= \beta_k + \eta \frac{\partial V_c(\alpha_k, \beta_k)}{\partial \beta} \end{aligned} \quad (5)$$

where $(\alpha_0, \beta_0)$ is an arbitrary starting strategy pair. Points on the boundary of the unit square (where at least one of $\alpha$ and $\beta$ is zero or one) have to be handled in a special manner, because the gradient may lead the players to an infeasible point outside the unit square. Therefore, for points on the boundary for which the gradient points outside the unit square, we redefine the gradient to be the projection of the true gradient onto the boundary. For ease of exposition, we do not change the notation in Equation 5 to reflect the projection of the gradient at the boundary, but the behavior there should be understood and is important to our analysis.

Note that the gradient ascent algorithm assumes a full information game — that is, both players know both game matrices, and can see the mixed strategy of their opponent at the previous step. (However, if only the actual previous move played is visible, we can define a stochastic gradient ascent algorithm.)

## 4 Gradient Ascent as Affine Dynamical System

If the row and column players were to play according to the gradient ascent algorithm of Equation 5, they would at iteration $k$ play the strategy pair $(\alpha_k, \beta_k)$, and receive expected payoffs $V_r(\alpha_k, \beta_k)$ and $V_c(\alpha_k, \beta_k)$ respectively. We are interested in the performance of the two players over time. In particular, we are interested in what happens to the strategy pair and payoff sequences over time. It is well-known in game theory that the *strategy pair* sequence produced by following a gradient ascent algorithm may never converge (Owen, 1995). In this paper we prove that the *average payoff* of both players always converges to that of some Nash pair, regardless of whether the strategy pair sequence itself converges or not. Note that this also means that if the strategy pair sequence does converge, it must converge to a Nash pair.

For the purposes of analysis, it is convenient to first consider the gradient ascent algorithm for the limiting case of infinitesimal step size ($\lim_{\eta \to 0}$); hereafter we will refer to this as the *IGA* (for Infinitesimal Gradient Ascent) algorithm. Subsequently we will show that the asymptotic convergence properties of IGA also hold in the more practical case of gradient ascent with decreasing finite step size. In IGA, the sequence of strategy pairs becomes a continuous trajectory in the unit square (though there are discontinuities at the boundaries of the unit square because of the projected gradient). The basic intuition behind our analysis comes from viewing the two players behaving according to IGA as a dynamical system in $\mathbb{R}^2$. In particular, as we show below the dynamics of the strategy pair trajectory is that of an affine dynamical system. This view does not take into account the constraint that the strategy pair has to lie in the unit square. This separation between the unconstrained dynamics and the constraints of the unit square will be useful throughout the rest of this paper.

Using Equations 3, 4 and 5 and an infinitesimal step size, it is easy to show that the *unconstrained* dynamics of the strategy pair as a function of time is defined by the following differential equation:

$$\begin{bmatrix} \frac{\partial \alpha}{\partial t} \\ \frac{\partial \beta}{\partial t} \end{bmatrix} = \begin{bmatrix} 0 & u \\ u' & 0 \end{bmatrix} \begin{bmatrix} \alpha \\ \beta \end{bmatrix} + \begin{bmatrix} -(r_{22} - r_{12}) \\ -(c_{22} - c_{21}) \end{bmatrix}. \quad (6)$$

We denote the off-diagonal matrix containing the terms $u$ and $u'$ in Equation 6 as $U$.

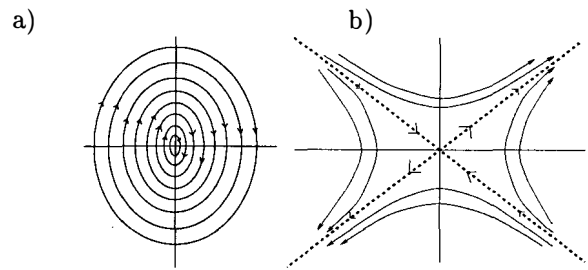

Figure 1: The general form of the dynamics: a) when $U$ has imaginary eigenvalues and b) when $U$ has real eigenvalues.

From dynamical systems theory (Reinhard, 1987), it is known that if the matrix $U$ is invertible (we handle the non-invertible case separately below), the unconstrained strategy pair trajectories can only take the two possible qualitative forms shown in Figure 1. Notice that these two dynamics are very different: the one in Figure 1a has a limit-cycle behavior, while the one in Figure 1b is divergent. Now depending on the exact values of $u$ and $u'$, the ellipses in Figure 1a can become narrower or wider, or even reverse the direction of the flow. Similarly, the angle between the dashed axes in Figure 1b and the direction of flow along the axes will depend on $u$ and $u'$. But these are the two general forms of unconstrained dynamics that are possible. In the next section we define the characteristics of general-sum games that determine whether the unconstrained dynamics is elliptical or divergent.

The *center* where the axes of the ellipses meet, or where the dashed-axes of the divergent dynamics meet, is the point at which the true gradient is zero. By setting the left hand side of Equation 6 to zero and solving for the unique center $(\alpha^*, \beta^*)$, we get:

$$(\alpha^*, \beta^*) = \left[ \frac{(c_{22} - c_{21})}{u'}, \frac{(r_{22} - r_{12})}{u} \right] \quad (7)$$

Note that the center is in general not at $(0, 0)$, and it may not even be inside the unit square.

## 5 Analysis of IGA

The following is our main result:

**Theorem 1** *(Nash convergence of IGA in iterated general-sum games) If in a two-person, two-action, iterated general-sum game, both players follow the IGA algorithm, their average payoffs will converge in the limit to the expected payoffs for some Nash equilibrium. This will happen in one of two ways: 1) the strategy pair trajectory will itself converge to a Nash pair, or 2) the strategy pair trajectory will not converge, but the average payoffs of the two players will nevertheless converge to the expected payoffs of some Nash pair.*

The proof of this theorem is complex and involves consideration of several special cases, and we present it in some detail below. But first we give some high-level intuition as to why the theorem is correct. First observe that if the strategy pair trajectory ever converges, it must be that it has reached a point with zero gradient (or zero projected gradient if the point is on the boundary of the unit square). It turns out that all such points must be Nash pairs because no improvement is possible for either player. More remarkably, it turns out that the average payoff of each ellipse in Figure 1a is exactly the expected payoff of the center (which is a point with zero gradient). But how is all this affected by the constraints of the unit square? Imagine taking a unit square and placing it anywhere in the plane of Figure 1a. The projected gradient along the boundary will be determined by which quadrant the boundary is in. We show that if there are some ellipses contained entirely in the unit square, the dynamics will converge to one such ellipse, and that if no ellipses are contained in the unit square (the center is outside the unit square), then the constrained dynamics must converge to a point. In either case, by the arguments above, the average payoff will become Nash. Similarly, imagine taking a unit square and placing it anywhere on the plane in Figure 1b. Given the gradient direction in each quadrant of the plane we show that the dynamics will converge to some corner of the unit square. Again, the average payoff will become Nash.

From dynamical systems (Reinhard, 1987) it can be shown that we only need to consider three mutually exclusive and exhaustive cases to complete a proof:

1. $U$ is not invertible. This will happen whenever $u$ or $u'$ or both are zero. Such a case can occur in team, zero-sum, and general-sum games. Examples of the dynamics in such a case are shown in Figure 2.

2. $U$ is invertible and its eigenvalues are purely imaginary. We can compute the eigenvalues by solving for $\lambda$ in the following equation:

$$\begin{bmatrix} 0 & u \\ u' & 0 \end{bmatrix} \begin{bmatrix} x \\ y \end{bmatrix} = \lambda \begin{bmatrix} x \\ y \end{bmatrix},$$

yielding $\lambda^2 = uu'$. Therefore we will get imaginary eigenvalues whenever $uu' < 0$. Such a case can occur in zero-sum and general-sum games but cannot happen in team games (because $u = u'$ and therefore $uu' \geq 0$). Two examples of the dynamics are shown in Figure 4.

3. $U$ is invertible and its eigenvalues are purely real. This will happen whenever $uu' > 0$. Such a case can occur in team and general-sum games but cannot happen in zero-sum games (because $u = -u'$ and therefore $uu' \leq 0$). Example dynamics are shown in Figure 6.

Theorem 1 is proved below by showing that Nash convergence holds in all three cases summarized above. But before we analyze these three cases in sequence in the next three subsections, we present a basic result common to all three cases that shows that if the $(\alpha(t), \beta(t))$ trajectory ever converges to a point, then that point must be a Nash pair.

**Lemma 2** *(Convergence of strategy pair implies convergence to Nash equilibrium) If, in following IGA, $\lim_{t \to \infty}(\alpha(t), \beta(t)) = (\alpha_c, \beta_c)$, then $(\alpha_c, \beta_c)$ is a Nash pair. In other words, if the strategy pair trajectory converges at all, it converges to a Nash pair.*

**Proof:** The strategy pair trajectory converges if and only if it reaches a point where the projected gradient is exactly zero. This can happen in two ways: 1) the point is the center $(\alpha^*, \beta^*)$, where by definition the gradient is zero (this can only happen if the center is in the unit square), or 2) the point is on the boundary and the projected gradient is zero. Either way, it means that from that point no local improvement is possible. For a proof by contradiction, assume that such a point is not a Nash pair. Then for at least one of the players, say the column player, there must be a unilateral change that increases their payoff. Let the improved point be $(\alpha_c, \beta_i)$. Then for all $\epsilon > 0$, $(\alpha_c, (1-\epsilon)\beta_c + \epsilon\beta_i)$ must also be an improvement. This follows from the linear dependence of $V_c(\alpha, \beta)$ on $\beta$ and the fact that the unit square is a convex region. Therefore the projected gradient at $\alpha_c, \beta_c$ must be non-zero. □

**Corollary 3** *If the center $(\alpha^*, \beta^*)$ is in the unit square it is a Nash pair.*

### 5.1  $U$ is not Invertible

**Lemma 4** *(Nash convergence when $U$ is not invertible) When the matrix $U$ is not invertible, the IGA algorithm leads the strategy pair trajectory to converge to a point on the boundary that is a Nash pair.*

**Proof:** First consider the case when exactly one of $u$ and $u'$ is zero. Without loss of generality assume that $u = 0$, i.e., $(r_{11} + r_{22}) = (r_{21} + r_{12})$. Then the gradient for the row player is constant (see Equation 3) and depending on its sign, the row player will converge to either $\alpha = 0$ or to $\alpha = 1$. Once the row player's strategy has converged, the gradient of the column player will also become constant (see Equation 4) and therefore it too will converge to an extreme value and therefore the joint strategy will converge to some corner. If both $u$ and $u'$ are zero, then both the gradients are constant and again we get convergence to a corner of the unit square.

In summary, if $U$ is not invertible the gradient algorithm will lead to convergence to some point on the boundary of the unit square, and hence from Lemma 2 will lead to convergence to a Nash pair of the game. □

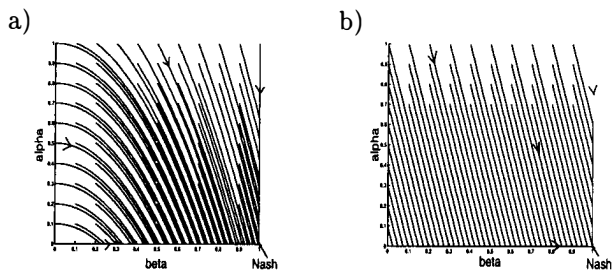

Figure 2: Example dynamics with $U$ not invertible. a) In this case $u$ is zero and $u'$ is not. b) Both $u$ and $u'$ are zero.

Figure 2 shows the dynamics for two general-sum games in which $U$ is not invertible. Figure 2a is for a case where $u = 0$ and $u' > 0$. The gradient for the row player is constant and points downwards. The gradient for the column player depends on $\alpha$, but once $\alpha$ converges to zero it point to the right and therefore from all starting points we get convergence to the bottom right corner. Figure 2b is for a case where both $u$ and $u'$ are zero. In this case both the gradients are constant and we get piecewise straight line dynamics.

### 5.2 $U$ has Purely Imaginary Eigenvalues

Purely imaginary eigenvalues occur when $uu' < 0$ in which case the two eigenvalues are $\sqrt{|u||u'|}i$ and $-\sqrt{|u||u'|}i$. It can be shown that in such a case the unconstrained dynamics are *elliptical* around axes determined by the eigenvectors of $U$ (Reinhard, 1987). See Figure 3a) for an illustration. There are two possible cases to consider: 1) $u > 0$ and $u' < 0$, and 2) $u < 0$ and $u' > 0$. However, without loss of generality we can consider only one case. When $u < 0$ and $u' > 0$,

$$\begin{bmatrix} 0 \\ \sqrt{\frac{|u'|}{|u|}} \end{bmatrix} + \begin{bmatrix} 1 \\ 0 \end{bmatrix} i$$

is a complex eigenvector corresponding to the eigenvalue $\sqrt{|u||u'|}i$, and

$$\begin{bmatrix} 0 \\ \sqrt{\frac{|u'|}{|u|}} \end{bmatrix} - \begin{bmatrix} 1 \\ 0 \end{bmatrix} i$$

is a complex eigenvector corresponding to the eigenvalue $-\sqrt{|u||u'|}i$. The axes of the ellipses in Figure 3a are determined by the real and imaginary parts of the two eigenvectors, that is, by the vectors

$$\begin{bmatrix} 0 \\ \sqrt{\frac{|u'|}{|u|}} \end{bmatrix} \text{ and } \begin{bmatrix} 1 \\ 0 \end{bmatrix}.$$

Note that these two vectors, and hence the axes of the ellipses, are always orthogonal to each other and parallel to the axes of the unit square. In the zero-sum case, because $|u| = |u'|$, they are also equal in size which means that the dynamics in the zero-sum case are circular (we merely observe this but do not use it hereafter). Note that the ellipses are centered at $(\alpha^*, \beta^*)$ and that the unit square may be anywhere in $\mathbb{R}^2$ and therefore the center can be outside the unit square.

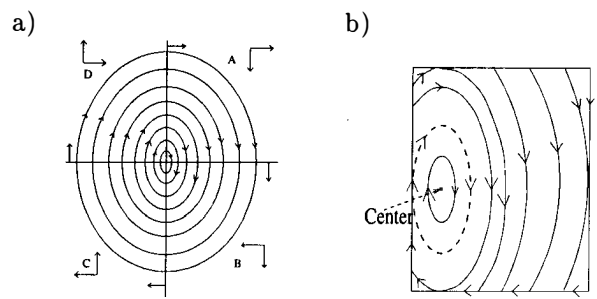

Figure 3: a) Unconstrained dynamics when $U$ has imaginary eigenvalues. b) The constrained dynamics when the center is in the unit square. In each case only some sample trajectories are shown.

We can solve the affine differential Equation 6 for the unconstrained dynamics of $\alpha$ and $\beta$ to get:

$$\alpha(t) = B\sqrt{u}\cos(\sqrt{uu'}t + \phi) + \alpha^* \quad (8)$$

and

$$\beta(t) = B\sqrt{u'}\sin(\sqrt{uu'}t + \phi) + \beta^* \quad (9)$$

where $B$ and $\phi$ are constants dependent on the initial $\alpha$ and $\beta$. These are the equations for the ellipses of Figure 3a. Note that if an ellipse happens to lie entirely inside the unit square then these equations also describe the constrained dynamics for any starting strategy pair that falls on that ellipse.

**Lemma 5** *(Nash Average Payoff for Ellipse entirely inside unit square)* For any initial strategy pair $(\alpha_0, \beta_0)$, if the trajectory given by Equations 8 and 9 lies entirely within the unit square, then the average payoffs along that trajectory are exactly the expected payoffs of a Nash pair.

**Proof:** Under the assumption that the ellipse lies entirely in the unit square, the average payoff for the row player can be computed by integrating the value obtained by the row-player in Equation 1 where the $\alpha$ and $\beta$ trajectories are those specified in equations 8 and 9. It can be shown that the integral of just the cosine term, just the sine term, and the product of the cosine and sine terms is exactly zero. This leaves just the terms containing $\alpha^*$ and $\beta^*$. Therefore, the average payoffs are exactly the expected payoffs of the center which by Corollary 3 is a Nash pair. □

Therefore when the center point of the ellipses is in the interior of the unit square, then all ellipses around it that lie entirely within the unit square have Nash payoffs.

Finally, we are ready to prove that when $U$ has imaginary eigenvalues, the average payoffs of the two players are always that of some Nash pair.

**Lemma 6** *(Nash Convergence in the case of imaginary eigenvalues) When the matrix $U$ has imaginary eigenvalues, the IGA algorithm either leads the strategy pairs to converge to a point on the boundary that is a Nash pair, or else the strategy pairs do not converge, but the average payoff of each player converges in the limit to that of some Nash pair.*

**Proof:** Consider again the unconstrained dynamics of Figure 3a. The four quadrants have the following general properties: in quadrant $A$ the gradient has a positive component in the down and right directions, in quadrant $B$ the gradient has a positive component in the down and left directions, in quadrant $C$ the gradient has a positive component in the up and left directions, and in quadrant $D$ the gradient has a positive component in the up and right directions. The direction of the gradient on the boundaries between the quadrants is also shown in Figure 3a. The important observation here is that the direction of the gradient in each quadrant is such that there is a clockwise cycle through the quadrants.

There are three possible cases to consider depending on the location of the center $(\alpha^*, \beta^*)$.

1. **Center is in the interior of the unit square.** First observe that all boundaries are tangent to some ellipse, and that at least one boundary is tangent to an ellipse that lies entirely within the unit square. For example, in Figure 3b the tangent ellipse to the left-side boundary lies wholly inside the unit square, while the other three boundarys' tangent ellipses are not contained in the unit square.

   If the initial strategy pair coincides with the center, we will get immediate convergence to a Nash equilibrium because the gradient there is zero. If the initial strategy pair is off the center point, then one of two things can happen: 1) either the ellipse that passes through the initial point does not intersect with the boundary, or 2) the ellipse that passes through the initial point intersects with a boundary. In the first case the dynamics will just follow the ellipse, and by Lemma 5 above, the average payoff for both players will be Nash, even though the strategy pairs themselves will not converge, but will follow the ellipse forever. In Figure 3b this will happen if the initial strategy pair is inside or on the dashed ellipse. In case 2) above, the strategy pair trajectory will hit a boundary, and then travel along it until it reaches a point at which the boundary is tangent to some ellipse that may or may not lie entirely in the unit square. If it does, then the trajectory will follow that ellipse thereafter. If it does not, then the trajectory will follow the tangent ellipse to the next boundary in the clockwise direction. This process will repeat until the boundary that has a tangent ellipse lying entirely within the unit square is reached. In Figure 3b, if the initial strategy pair starts anywhere outside the dashed ellipse, the dynamics will eventually follow the dashed ellipse. In all cases, from Lemma 5 we will get asymptotic convergence to the expected payoffs of some Nash pair.

2. **Center is on the Boundary.** Consider the case where the center is on the left-side boundary of the unit square. The first observation is that all points below the center on the left-side boundary will then have a projected gradient of zero. (Figure 3a shows that the gradient at such points will point left, and therefore outside the unit square and perpendicular to the left-side boundary.) The bottom boundary will have a projected gradient to the left. No matter where we start, we will either hit the bottom boundary, in which case we will get convergence to the lower left corner of the unit square, or we will hit the left boundary below the center, in which case again we will converge because of the zero projected gradient there. In either case, from Lemma 2 such points will be Nash pairs. By symmetry, a similar argument is easily constructed when the center is on some other boundary of the unit square.

3. **Center outside unit square.** There are two cases to consider: 1) the unit square lies entirely inside one quadrant, and 2) the unit square lies entirely inside two adjacent quadrants. No other case is possible. First consider the situation when the unit square lies entirely within quadrant $A$. Then the gradient at each point points down and right (see Figure 3a), and hence we will get convergence to the bottom right corner of the unit square. A similar argument is easily constructed when the unit square lies entirely within some other quadrant (yielding convergence to some other corner of the unit square).

Next consider the case where the unit square lies in quadrants $A$ and $D$. In quadrant $D$ the gradient points right and up, so either the trajectory will enter quadrant $A$ without hitting the top boundary, or it will hit the top boundary, in which case the projected gradient will be towards the right, and again it will enter quadrant $A$. In quadrant $A$ we will converge to the lower right corner of the unit square as above. A similar argument is easily constructed when the unit square lies within some other two adjacent quadrants (again yielding convergence to some corner of the unit square).

□

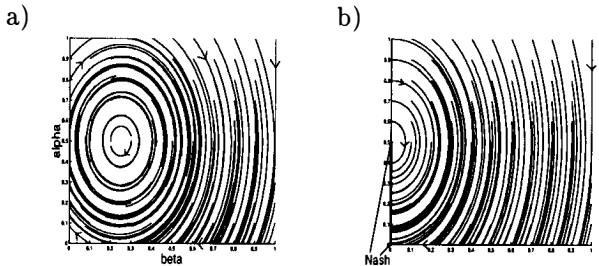

Figure 4: Example dynamics when $U$ has imaginary eigenvalues. a) The center is in the unit square. b) The center is on the left boundary of the unit square.

In Figure 4 we present examples of strategy pair trajectories for example problems whose $U$ matrices have imaginary eigenvalues. The left-hand figure is for a case where the center is contained in the unit square while the right-hand figure is for a case where the center is on the left-hand boundary of the unit square.

### 5.3 $U$ has Real Eigenvalues

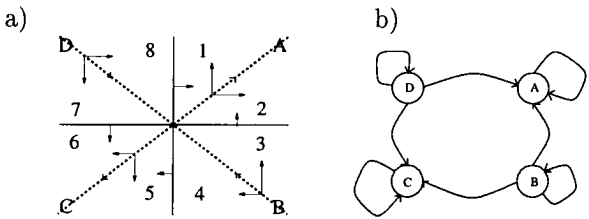

Figure 5: a) General characteristics of the unconstrained dynamics when $U$ has real eigenvalues. b) The possible transitions between the quadrants of the left-hand figure.

The unconstrained dynamics of a linear differential system with real eigenvalues are known to be divergent (Reinhard, 1987). See Figure 1b) for an illustration. Thus, without the constraints of the unit square, the strategy pair trajectory would diverge. Figure 5a shows the crucial general properties of the unconstrained dynamics. The center $(\alpha^*, \beta^*)$ is the point where the gradient is zero. Everywhere in quadrant $A$ the gradient has a positive component in the right direction and in the up direction; in quadrant $B$ the gradient has a positive component in the up direction and in the left direction; in quadrant $C$ the gradient has a positive component in the left and down directions; and in quadrant $D$ the gradient has a positive component in the right and down directions. At the boundary between quadrants $A$ and $D$ the gradient points left; at the boundary between $A$ and $B$ it points up; at the boundary between $B$ and $C$ it points to the right; and at the boundary between $C$ and $D$ is points down. The unit square that defines the feasible range of strategy pairs can be anywhere relative to the center. The eigenvectors corresponding to the two real eigenvalues, $\sqrt{uu'}$ and $-\sqrt{uu'}$ are

$$\left[\begin{array}{c}1\\ \sqrt{\frac{u'}{u}}\end{array}\right], \left[\begin{array}{c}1\\ -\sqrt{\frac{u'}{u}}\end{array}\right]$$

respectively (this is for the case that $u, u' > 0$; the analysis for the case that $u, u' < 0$ is analogous and omitted). The eigenvectors are represented in Figure 5a with dashed lines: one by drawing a line through the center and the point $(1, \sqrt{\frac{u'}{u}})$ and and the other by drawing a line through the center and the point $(1, -\sqrt{\frac{u'}{u}})$. Note that the general qualitative characteristics of the positive components of the gradient in the different quadrants do not depend on the eigenvectors. However, the eigenvectors are relevant to the detailed dynamics, as we will see in the examples below.

**Lemma 7** *(Nash convergence in the case of real eigenvalues) For the case of $U$ having real eigenvalues, the IGA algorithm leads the strategy pair trajectory to converge to a point on the boundary that is a Nash pair.*

**Proof:** Consider the graph of possible transitions between the quadrants in Figure 5b. From every point inside quadrant $A$, the gradient is such that the strategy pair will never leave that quadrant. Therefore if the strategy pair trajectory ever enters quadrant $A$, it will converge to the top right corner of the unit square. Similarly, from every point inside quadrant $C$, the gradient is such that the strategy pair will never leave that quadrant. Thus if the strategy pair trajectory ever enters quadrant $C$, it will converge to the lower left corner of the unit square. If the initial strategy pair is in quadrant $B$ or $D$, the dynamics is a bit more complex because it depends on the location of the unit square relative to the center. We consider only the case of quadrant $D$, for by symmetry a similar analysis will hold for quadrant $B$. Unless the unit square lies entirely in quadrant $D$, the strategy pair trajectory will enter quadrant $C$ or quadrant $A$, in which case we will get convergence to the associated corner as above. If the unit square is entirely within quadrant $D$, then the direction of the gradient in that quadrant will lead to convergence to the lower right corner of the unit square. Finally, if the right-hand boundary of the unit square is on the boundary between quadrants $D$ and $A$, then all the points on that boundary will have zero projected gradient, and any trajectory from $D$ hitting that boundary will converge there. Similarly, if the bottom boundary of the unit square is aligned with the boundary between quadrants $C$ and $D$, then any trajectory from $D$ hitting that boundary will converge there. From Lemma 2, if we ever get convergence to a strategy pair, it must be a Nash pair. □

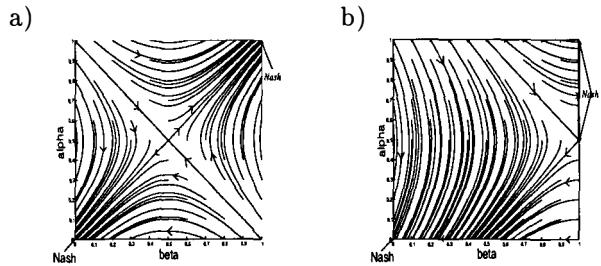

Figure 6: Example dynamics when $U$ has real eigenvalues. a) Center in the unit square. b) Center on the right boundary of the unit square.

In Figure 6 we present examples of strategy pair trajectories for example problems whose $U$ matrices have real eigenvalues. The left-hand figure has the center in the unit square; the right-hand figure has the center on the right-hand boundary of the unit square. The locations of the Nash points are shown.

## 6 Finite Decreasing Step Size

The above analysis and Theorem 1 have been about the IGA algorithm that assumes that both players are following the true gradient with infinitesimal step sizes. In practice, of course, the two players would use the gradient ascent algorithm of Equation 5 with a decreasing finite step size $\eta_k$ (where $k$ is the iteration number).

**Theorem 8** *The sequence of strategy pairs produced by both players following the gradient ascent algorithm of Equation 5 with a decreasing step size (several schedules will work, e.g., $\eta_k = \frac{1}{k^{2/3}}$) will satisfy one of the following two properties: 1) it will converge to a Nash pair, or 2) the strategy pair sequence will not converge, but the average payoff will converge in the limit to that of some Nash pair.*

**Proof:** (Sketch) Here we provide some intuition; the full proof is deferred to the full paper. Consider first the cases in which the IGA algorithm converges to a Nash pair. The proofs in such cases exploited only the direction of the gradient in the four quadrants around the center. These same proofs will extend without modification to the case of decreasing step sizes. The one case (with imaginary eigenvalues) in which the IGA algorithm does not converge to a point but instead converges to some ellipse fully contained in the unit square is more complex to handle. The basic intuition is that the strategy pairs cannot get "trapped" anywhere, and as the step size decreases, the dynamics of gradient ascent approaches the dynamics of IGA. □

## 7 Conclusion

Algorithms based on gradient ascent in expected payoff are natural candidates for adapting strategies in repeated games. In this work we analyzed the performance of such algorithms in the base case of two-person, two-action, iterated general-sum games, and showed that even though the strategies of the two players may not converge, the asymptotic average payoff of the two players always converges to the expected payoff of some Nash equilibrium. Our proof also provides some insight into special classes of games, such as zero-sum and team games.

In the future we will study the behavior of gradient ascent algorithms in complex multi-action and continuous action games in which the players use parameterized representations of strategies.